# Variational quantum classifiers via a programmable photonic microprocessor

Hexiang Lin[1*], Huihui Zhu[2*], Zan Tang[2†], Wei Luo[2], Wei Wang[2], Man-Wai Mak[2], Xudong Jiang[1†], Lip Ket Chin[3†], Leong Chuan Kwek[1†] and Ai Qun Liu[2†]

[1] *Quantum Science and Engineering Centre (QSec),Nanyang Technological University, Singapore 639798, Singapore*

[2] *Institute for Quantum Technology (IQT), The Hong Kong Polytechnic University, 11 Yuk Choi Rd, Hung Hom, Hong Kong*

[3] *Department of Electrical Engineering, City University of Hong Kong, Hong Kong, SAR 999077, China*

*[†]Corresponding Authors: zan-eee.tang@polyu.edu.hk; exdjiang@ntu.edu.sg; lkchin@cityu.edu.hk; cqtklc@nus.edu.sg; aiqun.liu@polyu.edu.hk*

*[*]These authors contributed equally to this work.*

## Abstract

Quantum computing holds promise across various fields, particularly with the advent of Noisy Intermediate-Scale Quantum (NISQ) devices, which can outperform classical supercomputers in specific tasks. However, challenges such as noise and limited qubit capabilities hinder its practical applications. Variational Quantum Algorithms (VQAs) offer a viable strategy to achieve quantum advantage by combining quantum and classical computing. Leveraging on VQAs, the performance of Variational Quantum Classifiers (VQCs) is competitive with many classical classifiers. This work implements a VQC using a silicon-based quantum photonic microprocessor and a classical computer, demonstrating its effectiveness in nonlinear binary and multi-classification tasks. An efficient gradient-free genetic algorithm is employed for training. The VQC's performance was evaluated on three synthetic binary classification tasks with square-, circular-, and sine-shape decision boundaries and a real-world multiclass Iris dataset. The accuracies on the three binary classification tasks were 87.5%, 92.5%, and 85.0%, respectively, and 90.8% on the real-world Iris dataset, highlighting the platform's potential to handle complex data patterns.

# 1. Introduction

Quantum computing holds the promise of revolutionizing various industrial sectors, including finance, logistics, electrical vehicles and sustainable engineering, through the achievement of quantum supremacy [1-3]. Significant efforts are underway to develop the essential hardware for quantum computers. Currently, noisy intermediate-scale quantum (NISQ) computers have demonstrated the ability to utilize more than 50 qubits, allowing them to outperform classical supercomputers in specific mathematical tasks [4-7]. However, the potential acceleration offered by quantum computers in practical applications is still largely unrealized due to issues related to noise and limitations in qubit performance [8-11]. Moreover, the widespread adoption of fault-tolerant quantum computers may still be years or even decades away.

One of the key technological challenges lies in how to utilize existing NISQ devices to achieve quantum advantage. In this context, variational quantum algorithms (VQAs) have emerged as a promising solution [8]. VQAs are optimization strategies that combine quantum and classical computing. They are executed on parametrized quantum platforms while the optimization of parameters is carried out by classical computers. This approach reduces the depth of quantum circuits, thereby mitigating noise and making VQAs a viable path to achieving quantum advantage using NISQ devices. Because VQAs provide a flexible framework suitable for various applications, they have been utilized across multiple domains of quantum computing, including eigenstate solving, dynamic quantum simulation, combinatorial optimization, and error correction and mitigation [12-24].

Variational Quantum Classifiers (VQCs), a derivative of VQAs, are specifically designed to tackle classification problems commonly encountered in machine learning. VQCs have shown superior performance compared to classical machine learning algorithms in various tasks, such as image classification, fraud detection, and natural language processing [25-29]. However, a significant challenge for VQCs is the need to access quantum computers, which substantially increases the cost of training. Due to this constraint, much of the research has focused on theoretical aspects, exploring the potential advantages and applications of VQCs, while discussions on experimental validation and demonstrations on real quantum platforms are relatively limited. This situation makes it difficult for researchers to evaluate the practical effectiveness of VQCs, as there might be discrepancies between theoretical models and their actual implementations. Therefore, effectively demonstrating the capabilities of VQCs on physical platforms remains a crucial area of study.

In this study, we demonstrate a VQC using a silicon-based quantum photonic microprocessor with a classical computer for applications in nonlinear binary classification tasks and more complex multi-classification. During the training process, we utilize an efficient, gradient-free method called the genetic algorithm, which was experimentally demonstrated on the hybrid model. This online approach is simpler to implement compared to traditional gradient-based algorithms. We then employ the quantum photonic microprocessor combined with the gradient-free genetic algorithm to conduct proof-of-concept demonstrations for classifying linearly inseparable data with square, circular, and sine decision boundaries and Iris species in a real-world multi-class dataset. The results show excellent alignment between the predicted labels and the actual labels for the data with square and circular decision boundaries. These findings demonstrate the potential and versatility of our platform, highlighting its capability to effectively handle complex data patterns.

## 2. Variational quantum neural network model

In this work, we use a hybrid variational quantum neural network to perform several classification tasks. The operational procedure is illustrated in **Figure 1**. The input, unclassified samples are represented as $\vec{x}_i$, where $\vec{x}_i$ denotes the *n*-dimensional vector for the $i^{\text{th}}$ sample. These data are fed into the blue hybrid loop, which serves as the core framework of the system. The hybrid loop comprises three fundamental elements of the VQA: the ansatz, the cost function, and the optimizer. In this loop, the input data $\vec{x}_i$ are loaded onto the amplitude part of a pure state $|\psi_0\rangle_i$ by adjusting the transformation of the multiple Mach-Zehnder Interferometers (MZIs), which composed of beam splitters and phase shifters. The ansatz is implemented using a quantum photonic chip composed of a network of MZIs. This evolution process can be described using an evolution operator $\hat{U}(\vec{\theta}^j, \vec{\varphi}^j) = \hat{U}(\theta_6^j, \varphi_6^j) \ldots \hat{U}(\theta_k^j, \varphi_k^j) \ldots \hat{U}(\theta_1^j, \varphi_1^j)$, whereby the parameters $\vec{\theta}^j$ and $\vec{\varphi}^j$ can be optimized and controlled by adjusting the voltage on the phase shifter. The structure of the chip and the specific expression of the operator will be explained in detail in the next section.

After the evolution on the quantum photonic chip, the initial state evolves into the final state, expressed as $|\varphi_f\rangle_i = \hat{U}(\vec{\theta}^j, \vec{\varphi}^j)|\varphi_x\rangle_i$. In our approach, the cost function using the $l_2$-norm is defined as

$$C(\vec{\theta}^j, \vec{\varphi}^j) = \sum_i |||\varphi_f\rangle_i - y_i||^2, \tag{1}$$

where $|\varphi_f\rangle_i$ represents the output light intensity vector of the $i^{th}$ sample, and $y_i$ is the class label for the $i^{th}$ sample. By collecting data from single-photon detector measurements, we can calculate the cost function. Theoretically, the cost function defines a hypersurface, as depicted in **Figure 1**. Our goal is to find a good minimum point on this surface. Consequently, the entire training process involves continuously optimizing the parameters to minimize the cost.

The classical computer serves as the optimizer for the variational quantum neural network. It employs a genetic algorithm to update the parameters of the quantum ansatz, using the cost function as the fitness function. In genetic algorithms, a solution to an optimization problem is referred to as an "individual", which can be a string of characters or numbers. The collection of all possible solutions is known as the "population". In our context, an individual consists of a set of parameters $P^j = \{\varphi_k^j, \theta_k^j, k = 1,2,3,4,5,6\}$ that define the transformation matrix. Here, $k$ represents the position of a specific Mach-Zehnder Interferometer (MZI), and $j$ denotes the number of training generations. The genetic algorithm consists of four main entities: randomly generated initial possible solutions (individuals), selection of the most suitable solutions (next generation), random bit value changes (mutations), and partial exchanges between different individuals (crossover). Further details can be found in **Appendix A**. The optimization training process concludes when the training reaches 100 generations in our experiment. Once this process is complete, we can obtain the set of optimized parameters for the quantum chip. Ultimately, the hybrid variational quantum neural network model fully implements the classical data encoding, the training of the quantum circuit, and the calculation of the cost function.

## 3. Design and experimental setup

The experimental setup, depicted in **Figure 2(a)**, consists of three main components: a laser source, a quantum photonic circuit, and a single-photon detector. The process first directs laser light through a wavelength division multiplexing system to achieve a wavelength of 1550 nm. This light is then amplified using an Erbium-doped optical fiber amplifier, and a filter further purifies the signal to 1500 nm to reduce amplified noises. Prior to entering the quantum chip, a polarization controller is used to adjust the input light mode, which is sensitive to the fiber's orientation. Each time the fiber is repositioned, we recalibrate the polarization controller to ensure that the transmission mode aligns with the input light mode. Finally, the light is vertically coupled into the chip through a single fiber, where it is converted to single-photon signals by the on-chip spiral waveguides. The resulting output photons are collected by a fiber array and transmitted to the superconducting single-photon detector.

In our quantum ansatz, photon pairs are initially generated in the spiral waveguides through a spontaneous four-wave mixing process (**Figure 2(b)**). Asymmetric Mach-Zehnder Interferometers (AMZIs) are then used to split the produced photon pairs. One of the photon pairs enters three interferometers to create an initial qudit state $|\psi_0\rangle$ with modifiable amplitude components for input data loading. Next, the encoded input state is fed into a four-mode interferometer network to facilitate the evolution process $U(\theta, \varphi)$. This process involves a variational circuit consisting of six MZI units. Each MZI unit is composed of two phase shifters and two 50:50 beam splitters. The matrix representation for the phase shifter is denoted by $U_{PS}(\xi) = \begin{pmatrix} e^{i\xi} & 1 \\ 1 & 1 \end{pmatrix}$, where $\xi$ indicates the phase difference between the two paths. The 50:50 beam splitters distribute the light equally, and their matrix representation is given by $U_{BS}\left(\frac{\pi}{2}\right) = \begin{pmatrix} \cos\frac{\pi}{2} & i\sin\frac{\pi}{2} \\ i\sin\frac{\pi}{2} & -\cos\frac{\pi}{2} \end{pmatrix}$. The matrix for a single MZI can be expressed as $U_{MZI} = U_{BS}(\frac{\pi}{2})U_{PS}(\theta)U_{BS}(\frac{\pi}{2})U_{PS}(\varphi) = ie^{\frac{i\theta}{2}} \begin{pmatrix} e^{i\varphi}\sin\left(\frac{\theta}{2}\right) & \cos\left(\frac{\theta}{2}\right) \\ e^{i\varphi}\cos\left(\frac{\theta}{2}\right) & -\sin\left(\frac{\theta}{2}\right) \end{pmatrix}$. Here, $\varphi$ determines the relative phase between the two input ports, and $\theta$ indicates the transmission rate of the MZI. The programmable aspect of the system is achieved by adjusting the current $I$ supplied to the titanium nitride heater, which manipulates the refractive index. The change in relative phase is proportional to temperature and, consequently, to the heating power ($\Delta n \propto I^2$). Finally, the output photon from the variational circuit and the heralding photon, which serves as the trigger signal, are collected and measured using twofold photon detection. A detailed analysis of the matrix formulation of the four-mode circuit can be found in **Appendix B**.

In this process, all programmable phase shifters and detected photons can be controlled using a classical computer, which also manages the updating process of the genetic algorithm. The chip is packaged optically and electrically, with electrodes mounted to a printed circuit board via wire bonding, as demonstrated in **Figure 2(c)**. A calibration curve for the phase shifter is displayed in **Figure 2(d)**, and the measured Hong-Ou-Mandel (HOM) interference of the produced photon pair is shown in **Figure 2(e)**, which has a visibility of $V_{HOM} = 0.80 \pm 0.05$. With the demonstrated programmability and quantum characteristics of the photonic microprocessor, we will use this quantum chip to investigate several classification problems.

## 4. Results and discussion

### 4.1. Classification of linearly inseparable datasets

In this study, we evaluate the nonlinear classification capabilities of our quantum microprocessor using three types of samples with different nonlinear decision boundaries: square, circular, and sine. The quantum nonlinear classifier consists of three main components: data loading, unitary transformation, and readout. During data loading, the input data is introduced into the circuit through specific quantum gates, which then pass through a series of additional quantum gates. The parameters of these gates are analogous to the learnable weights in a neural network. After processing by the data loader, the signal is detected by a photon detector. In the data loading process, a four-dimensional input state $|\psi_0\rangle$ is encoded as $|\psi_0\rangle = (\cos(x_1)\cos(x_2), \cos(x_1)\sin(x_2), \sin(x_1)\cos(x_2), \sin(x_1)\sin(x_2))^T$, in which $(x_1, x_2)$ is the coordinate of a point (a sample) in these datasets. The variational circuit is trained with six complex-valued parameters to map the input data to two output modes. The highest photon measurement outcome is used to determine the label of each sample.

The variation in prediction accuracy with respect to training generations is illustrated in **Figures 3(a)-(c)**. In these figures, the blue dotted line represents the average accuracies, while the red dotted line indicates the maximum accuracies. We use 600 samples for the training process, with 50 individuals per generation to ensure genetic stability. Detailed parameters are seen in **Appendix C**. Generally, both average and maximum accuracy improve as training progresses. The maximum training generation limit is set at 100, and the classification accuracy approaches approximately 90% after 60 generations. Note that fluctuations occur during the training process, particularly evident in **Figure 3(b)**. However, these fluctuations eventually become stabilized, a normal characteristic of the convergence dynamics. In earlier generations, significant changes in fitness can be observed, while later generations tend to stabilize as the population converges to optimal or near-optimal solutions. The classification results are presented in **Figures 3(d)-(f)**. The background color indicates the ground truth, whereas the color of the dots represents the predicted labels of the testing samples. The genetic algorithm is implemented using the built-in functions of MATLAB. The simulated classification accuracies are 95.8% for square samples, 95.5% for circular samples, and 94.4% for sine function samples. These results show that the predicted labels for the square and circular samples closely align with the decision boundary of the ground truth, while the sine function samples exhibit slight discrepancies due to its more complex boundary. These simulation results demonstrate that our simple quantum variational circuit effectively achieves nonlinear boundary classifications.

Additionally, we conduct experimental demonstrations with hardware training using a variational quantum circuit chip. A total of 100 samples are utilized, with 20 individuals selected in each training generation. The close matching between the colored dots and the background color in **Figure 4** indicates high classification accuracy in the experiments, with the highest accuracies of 87.5%, 92.5%, and 85.0% for square, circular, and sine curve samples, respectively. The decrease in accuracy compared to the simulation results is attributed to the reduced number of individuals in each training generation, representing a trade-off between classification accuracy and the time required for hardware training. The 50 training samples per class cannot well shape the nonlinear decision region. Nevertheless, as a demonstration of the variational quantum classification algorithm with online training, the experimental results are sufficient to show that this quantum circuit can effectively perform nonlinear classification.

**4.2. Classification on real-world dataset Iris**

The Iris flower dataset, a typical classification dataset, is used to showcase the VQA and the hardware performance of our chip. This dataset contains four input parameters: the lengths and widths of the sepals and petals of various Iris flowers. The task is to classify each flower into one of three subspecies: setosa, versicolor, or virginica. To perform this classification task, we utilize a variational quantum circuit structured as a one-layer neural network. The four input features are encoded in the initial state using a method similar to that described in the previous task (details can be found in **Appendix D**). The dataset consists of 150 instances, which is only a quarter the size of the previous dataset. To ensure genetic stability during the training process, we use a population size of 150 in our simulations and 50 in our hardware experiments. We allocate 80% of the dataset for training and reserve the remaining 20% for testing.

The software simulation and hardware realization for training and classifying Iris flowers are illustrated in **Figure 5**. The training and updating processes in the simulation are depicted in **Figures 5(a)** and **(b)**, which involve 150 individuals per generation. It is evident that the maximum accuracy values increase, while the fitness values decrease throughout the training generations, ultimately converging to stable values around the 80th generation. The maximum accuracy achieved in the simulation is 90.8%, indicating that the generalized model fits well with unseen data. Furthermore, the experimental training process (on-chip) is presented in **Figures 5(c)** and **(d)**. The convergence speed in this case is slower compared to the simulation results, likely due to the smaller number of individuals in each generation during the hardware training. However, the final training accuracy remains similar to that of the simulation, demonstrating the algorithm's adaptability to hardware and its robustness against perturbations. After training, the performance of the trained hardware circuit is evaluated on the Iris flower classification testing dataset, as shown in **Figures 5(e)** and **(f)**. A total of 120 training samples and 30 testing samples are experimentally assessed, with only seven and two misclassifications highlighted by blue and red circles in **Figure 5(e)** for the training and testing samples, respectively. The confusion matrix, which shows a high classification accuracy of 93.3% across 30 testing images, is illustrated in **Figure 5(f)**. These results confirm that our variational quantum circuit possesses the capability to perform practical machine-learning tasks with high accuracy.

## Discussion and Conclusion

In this study, we implement a VQC using a silicon-based quantum photonic chip and a classical computer and show its capacity in learning-based classification of samples with nonlinear decision boundaries, as well as classification of the real-world Iris dataset. The learning capacity and prediction performance shown by our implemented VQC in the experiments are promising, demonstrating high accuracy with a minimal number of training parameters. A multi-layer configuration can be accomplished by cascading multiple unitary transformation matrices of the same size, which could improve classification performance as the number of trainable parameters increases, enabling the model to capture more complex relationships within the data (see **Figure 6**). Each additional layer contributes to the increase of the expressive power, allowing the VQC to better approximate the decision boundaries necessary for more accurate classification. Thus, investigating the impact of this architectural change could lead to significant improvements in performance across various datasets. Moreover, our hardware platform facilitates resource recycling by encoding the measurement results from one layer into the input state of the next layer, which can be continued through the final layer. This recycling mechanism not only optimizes the use of the integrated photonic circuit but also preserves the integrity of quantum information throughout the classification process.

In addition to the genetic algorithm employed in our experiments, other training methods, such as classical gradient descent and quantum gradient descent techniques, can also be applied to our parameterized circuit for classification tasks. Notably, the genetic training method exhibited noise-free operation with high training accuracy on our hardware platform. Future research should investigate the resilience of various training methods to noise, especially as circuit depth increases. Evaluating how these different algorithms perform under noisy conditions will be crucial for determining the practicality of VQCs in real-world applications, where noise is an inherent challenge.

In conclusion, this study demonstrates the viability of VQCs implemented on a silicon-based quantum photonic chip with a classical computer, successfully addressing nonlinear binary and multi-classification tasks. By employing a gradient-free genetic algorithm for training, we achieved efficient modeling, overcoming the limitations associated with traditional gradient-based methods. The results show a strong alignment between predicted and actual labels for square and circular datasets, with the sine function samples reaching an accuracy close to 90%. These findings highlight the potential of our quantum photonic microprocessor to effectively manage complex data patterns, paving the way for further exploration of VQCs in practical applications. As we continue to leverage the capabilities of NISQ devices, this work contributes to the growing body of evidence supporting the application of quantum computing in machine learning, while also highlighting the need for ongoing experimental validation to bridge the gap between theory and practice.

## Appendix A: Classical optimization algorithm

A Genetic Algorithm is a search algorithm inspired by the process of evolution. The solutions generated by a genetic algorithm are represented as a series of numbers called "chromosomes". For example, a chromosome can be denoted as $S = 10101101$. Each digit within this string is referred to as a "gene". There are three main operations in a genetic algorithm: selection, crossover, and mutation. Initially, a population is randomly generated, which involves determining both the number of individuals and the genetic characteristics of those individuals. Individuals are then selected based on a self-defined fitness function, whereby those with higher fitness values have a greater probability of being chosen. Crossover is the process where two selected individuals partially exchange their chromosomes. This involves randomly selecting two individuals from the population and combining their chromosome segments to inherit the best traits from the parent chromosomes, thereby generating new, potentially superior individuals. For an example of single-point crossover, if we have the individuals 10101 and 11001, the result would be 10 001 and 11101. The mutation operation randomly flips the values of certain bits in the chromosome (usually single bits) to prevent the algorithm from getting stuck in local minima.

## Appendix B: Matrix operation on the integrated chip

Our MZI network is operated in a four-model circuit, and its mathematical representation should be the direct sum of its two-mode form and the corresponding identity matrix depending on its position in the network. For example, the mathematical representation of

MZI 1 is $\widehat{U}(\theta_1^i, \varphi_1^i) = \begin{pmatrix} e^{i\varphi}\sin\left(\frac{\theta}{2}\right) & \cos\left(\frac{\theta}{2}\right) & 0 & 0 \\ e^{i\varphi}\cos\left(\frac{\theta}{2}\right) & -\sin\left(\frac{\theta}{2}\right) & 0 & 0 \\ 0 & 0 & 1 & 0 \\ 0 & 0 & 0 & 1 \end{pmatrix}$,

MZI 2 is $\widehat{U}(\theta_2^i, \varphi_2^i) = \begin{pmatrix} 1 & 0 & 0 & 0 \\ 0 & 1 & 0 & 0 \\ 0 & 0 & e^{i\varphi}\sin\left(\frac{\theta}{2}\right) & \cos\left(\frac{\theta}{2}\right) \\ 0 & 0 & e^{i\varphi}\cos\left(\frac{\theta}{2}\right) & -\sin\left(\frac{\theta}{2}\right) \end{pmatrix}$ and

MZI 3 is $\widehat{U}(\theta_3^i, \varphi_3^i) = \begin{pmatrix} 1 & 0 & 0 & 0 \\ 0 & e^{i\varphi}\sin\left(\frac{\theta}{2}\right) & \cos\left(\frac{\theta}{2}\right) & 0 \\ 0 & e^{i\varphi}\cos\left(\frac{\theta}{2}\right) & -\sin\left(\frac{\theta}{2}\right) & 0 \\ 0 & 0 & 0 & 1 \end{pmatrix}$.

The expression for MZI 4, 5, and 6 is similar, differing only in the parameters. This corresponds to the expression $\widehat{U}(\vec{\theta}^i, \vec{\varphi}^i) = \widehat{U}(\theta_6^i, \varphi_6^i) \ldots \widehat{U}(\theta_k^i, \varphi_k^i) \ldots \widehat{U}(\theta_1^i, \varphi_1^i)$ where each $\widehat{U}(\theta_k^i, \varphi_k^i)$ is represented as above depending on the position of the MZI.

## Appendix C: Training process in nonlinear boundary classification

In the nonlinear boundary classification task, we employ a genetic algorithm to iteratively update the set of potential solutions based on their fitness function values. To maintain genetic stability, we have established a maximum training generation limit of 100, with each generation consisting of 50 individuals. The crossover fractions are uniformly set at 0.3 for square, circular, and sine boundaries, facilitating a balanced exploration of the solution space. Additionally, the migration fractions are set to 0.5, 0.7, and 0.5 for the square, circular, and sine boundaries, respectively, allowing for a controlled exchange of genetic material between subpopulations. This structured approach aims to enhance the algorithm's efficacy in identifying optimal solutions while ensuring diversity within the population.

## Appendix D: Data encoding in Iris classification

Due to the limitation in the number of beam splitters, we reconstruct the original four-dimensional data into three dimensions to encode the transmittance of the three beam splitters. The relationship between two kinds of data is $y_1 = x_1, y_2 = x_2 + x_3, y_3 = x_3 + x_4$. The four-dimensional amplitude vectors is represented by $|\varphi_0\rangle = (\cos(y_1) \cdot \cos(y_2), \cos(y_1) \cdot \sin(y_2), \sin(y_1) \cdot \cos(y_3), \sin(y_1) \cdot \sin(y_3))^T$.

**Funding.** This work was supported by The Hong Kong Polytechnic University Start-up Fund and pair research fund for Institute of quantum technology.